\documentclass[12pt,a4paper]{article}
\usepackage{amsmath}
\usepackage{amsthm}
\usepackage{amsfonts}
\usepackage{multicol}
\usepackage[dvips]{graphicx}
\usepackage{fixltx2e}

\newcommand{\danger}[1]{\textbf{#1}}

\input{epsf}             

\newtheorem{definition}{Definition}
\newtheorem{theorem}{Theorem}

\begin{document}

\title{\danger{Measurements and Information in Spin Foam Models}}
\author{\centerline{\danger{J. Manuel Garc\'\i a-Islas \footnote{
e-mail: jmgislas@leibniz.iimas.unam.mx}}}  \\
Departamento de F\'\i sica Matem\'atica \\
Instituto de Investigaciones en Matem\'aticas Aplicadas y en Sistemas \\ 
Universidad Nacional Aut\'onoma de M\'exico, UNAM \\
A. Postal 20-726, 01000, M\'exico DF, M\'exico\\}

\maketitle

\begin{abstract}
In the three dimensional spin foam model of quantum gravity with a cosmological constant, 
there exist a set of observables associated to spin network graphs. A set of probabilities 
is calculated from these observables, and hence the associated Shannon entropy can be defined. We present the Shannon entropy associated to these observables
and find some interesting bounded inequalities. The problem relates mathematical measurements, entropy and 
information theory in a simple way which we explain.     
\end{abstract}

\section{Introduction}

Entropy is a subject studied in information theory and other areas of physics. The entropy of measurements of an 
experiment with outcomes with certain probabilities. The one we study here is related to mathematical
measurements. What we mean will be understood along the paper, however, basically this means that the entropy
we study here has a lot of mathematical sense as it is related to the geometry of space-time.

In the present paper we depart from geometrical measurements in three dimensional quantum gravity associated to spin network observables, 
in the way it was introduced in \cite{jwb} and which were later used to develop 
an idea for dealing with entropy in the spin foam model description of quantum gravity in \cite{gi1}, \cite{gi2}, 
\cite{gi3}.

The corresponding measurements, which meaning will be understood later on, of spin network graphs have certain probabilities assigned. These probabilities are computed by a specific
partition function which we will recall. Once these probabilities are obtained
the Shannon entropy of each spin network can be computed in general. 
However, to obtain specific values of the Shannon entropy of each spin network graph 
one would need a computing program. As our task is only in a theoretical way we just present
how the Shannon entropy of each spin network is bounded. The interesting thing is that it is bounded by some very well known values used in conformal field theory; the logarithm of
the dimension
of the space of conformal blocks.

We divide this paper as follows: In section 2 we recall the notion of observables
and their assigned probabilities
in three dimensional quantum gravity with cosmological constant
known mathematically as the Turaev-Viro model. We also explain
how these probabilities are related to mathematical meausurements.

In section 3 we present the Shannon entropy of the observables. First we give 
the already known definition of the Shannon entropy and then we present some examples
for calculating it. We show how in very simple examples the Shannon entropy is bounded
by the logarithm of the dimension of the space of conformal blocks.
This section is what 
is new in our approach of defining entropy in spin foam models. 

In section 4 we describe a way to generalise what we present in section 3.

\section{Measurements}

Generally speaking the partition fucntion of spin foam models is as follows:

A spin foam partition function is based on a triangulation $\triangle$ of the space-time $M$ manifold
or on the dual two-complex $\mathcal{C}$ of the triangulation and 
is given by

\begin{equation}
Z_{\mathcal{C}}=  \sum_{S} \prod_{faces} A(j_f) \prod_{vertices} A(v)
\end{equation}
where the sum is carried over a set of states
$S$ given by the colouring of faces by irreducible representations of $SU(2)$ and of
intertwiners associated to the edges, that is, coloured by a colour which
belong to the infinite set $j_f \in \{ 0, 1/2, 1, 3/2,..(r-2)/2.... \}$.
The sum is over the product of amplitudes associated to the faces and to the
vertices of the two-complex. 
$A(j_f)$ is an amplitude associated to the faces of the two-complex given by the 
dimension of the representation which colours the corresponding face and
$A(v)$ is a function which depends on the colours of the faces adjacent to the 
vertex and on the intertwiners of the edges adjacent to the vertex. The function
depends on the particular spin foam model we are considering. 
Between the most important spin foam models which have been introduced are 
\cite{bc}, \cite{epr}, \cite{ls}, \cite{fk}.

A special case of spin foam models of three dimensional quantum gravity with cosmological
constant is the Turaev-Viro model \cite{tv}.  
The definition of the model is as follows.\footnote{For our convenience we use integers instead of
half-integers. The results are completely equivalent.} 

Given a triangulated three dimensional space-time manifold $M$ 
and a finite set of indices $L= \{ 0,1,2,...(r-2) \}$, where $r\geq3$,
we define a state as a
function from the set of edges $S: \{edges\} \longrightarrow L$. A
state is called admissible if at each face of the triangulation, the
labels $(i,j,k)$ of the corresponding edges satisfy the following
identities:

\begin{eqnarray}
0 \leq i , j , k \leq r-2 \nonumber \\
i \leq j+k , \  j \leq i+k , \  k \leq i+j \nonumber \\
i+j+k \in 2\mathbf{Z} \nonumber \\
i+j+k \leq 2(r-2)
\end{eqnarray}
The spin foam partition function is given by

\begin{equation}
Z_M=  N^{-v} \sum_{S} \prod_{edges} A(j_e) \prod_{tetrahadra} A(\bigtriangleup)
\end{equation}
where we have a product of the amplitudes $A(j_e)$ associated to the edges and of
the amplitudes associated to the tetrahedra $A(\bigtriangleup)$.
The sum is carried over the set of all admissible states
$S$.
The amplitude associated to the edges is given by

\begin{equation}
A(j_e): = \dim_{q}  j_e =  (-1)^{j_e} \bigg( \frac{\sin \frac{\pi}{r}(j_e+1)}{\sin \frac{\pi}{r}} \bigg) 
\end{equation}
 and the amplitude associated to the tetrahedra   
is given by the quantum $6j$ symbol of the six labels of its edges.
 $v$ is the number of vertices of the triangulation and
$N= \sum_{j_e=0}^{r-2} \ (\dim_{q} j_e)^{2}$.  
The state sum (3) is the partition function of quantum gravity in three dimensions with a 
cosmological constant. Mathematically it gives an invariant of the three dimensional manifold $M$. 

Let us now describe the set of observables we are going to deal with, since we will define 
their entropy.

In \cite{gi4}, observables for the
Turaev-Viro spin foam model were introduced. The idea was constructed in the spirit of  
\cite{jwb}. We recall it here in a more formal way than previous references.

\begin{definition}
{Let $M$ be the triangulated three dimensional manifold.
An observable is a subset of $n$ edges of the triangulation $\mathcal{O} \subset M$.
Let $j_1, j_2, ... , j_n$ be labels at the $n$ edges
of $\mathcal{O}$} 
\end{definition}
Let an observable $\mathcal{O}$ with $n$ edges labelled $j_1, j_2, ... , j_n$.   
Consider the partition function

\begin{equation}
Z_{\mathcal{O}}(j_1,j_2,...,j_n) = \sum_{S \mid
\mathcal{O}} \prod_{edges} \dim_{q} j \prod_{tetrahedra} \{ 6j \}
\end{equation}
This is similar to the state sum formula (3), with the only
difference that now we are not summing over the spins $j_1, j_2, ... , j_n$ which label
the graph observable $\mathcal{O}$. 

\begin{definition}{The probability of the observable
$\mathcal{O}$ with edges labelled $j_1,j_2,...j_n$, is given by

\begin{equation}
P_{\mathcal{O}} =
\frac{Z_{\mathcal{O}}(j_1,j_2,...j_n)}{Z_M}
\end{equation}
This probability is of course well defined when $Z_M \neq 0$.}
\end{definition}
Let us clarify something about the probabilities of the observables. First of all, the numbers are real
and strictly positive. This is because we only have admissible colourings. It is also
true that, $\sum P_{\mathcal{O}}=1$, for all the different values of $(j_1,j_2,...j_n)$.

The probabilities we are defining here are the ones we are referring to as measurements. The question
is, what is measured? The answer to this question depends on the observable graph and 
therefore is understood with examples.

Some examples of observables and of these probabilities were considered 
and formally computed in \cite{gi4}. 

\bigskip

\danger{Example 1:}
If the observable $\mathcal{O}$
consists of a single edge labelled $j$, it was formally shown in \cite{gi4} that

\begin{figure}[h]
\begin{center}
\includegraphics[width=0.3\textwidth]{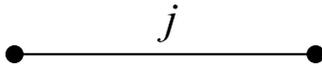}
\caption{Edge observable labelled $j$}
\end{center}
\end{figure}

\begin{equation}
P_j= \frac{( \dim_{q} j)^{2}}{N}
\end{equation}
For the edge observable graph we observe that we have a set of probabilities 
where $j$ goes from $0$ to $(r-2)$. These set of probabilities is what is measured. But, 
what is the interpretation of these measures? The interpretation is geometrical in
nature and it is described in \cite{jwb}. When $r \rightarrow \infty$, although the limit is
degenerate, the probability is interpreted to be related to the area of a 2-sphere 
of radius $(j+1)/2$. The 2-sphere is embbeded in $S^3$ of radius $r/2\pi$. 
We can therefore say that the interpretation of the measurements are geometrical.

The area of a 2-sphere of radius $(j+1)/2$ in $S^3$ of radius $r/2\pi$ 
in terms of the probabilities is given by 

\begin{equation}
(\text{area})_j= 4 \pi N \sin^2 \frac{\pi}{r} \ P_j
\end{equation}
The different areas are proportional to the probabilities.

What is measured is different areas of two spheres of radii which go
from $1/2$ when $j=0$ to $(r-1)/2$ when $j=(r-2)$.

\bigskip

\danger{Example 2:}
If the observable $\mathcal{O}$ consists of a triangle whose edges
are labelled $i$, $j$ and $k$, it was formally shown in \cite{gi4} that

\begin{figure}[h]
\begin{center}
\includegraphics[width=0.3\textwidth]{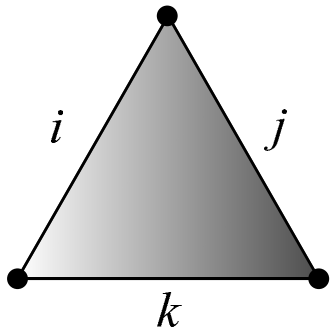}
\caption{Triangle observable with edges labelled $i$, $j$, $k$}
\end{center}
\end{figure}

\begin{equation}
P_{(i,j,k)} = \frac{\dim_{q} i \dim_{q} j \dim_{q} k}{N^2} \ N_{i,j,k}
\end{equation}
where $N_{i,j,k}$ is the dimension of the space of intertwiners,
i.e. equal to 1 if the spins are admissible and 0 otherwise. 
But since we are considering admissible colourings from the definition we always have that
the probabilities are never zero.

The interpretation of what is measured here 
for the triangle observable is as follows: 
Given a triangle observable whose edges are labelled
$i$, $j$, $k$, the probability that the distances between its vertices are $(i+1)/2$, 
$(j+1)/2$, $(k+1)/2$ is given by formula $(9)$.

\bigskip

In the next section we will define and study the Shannon entropy of these examples 
and will consider some generalisations in section 4.

\section{Shannon entropy and information}   
 
In this section we define and study the Shannon entropy of the probabilities of the observable spin network graphs.
We will show some bounded properties of the previous examples of last section.  
Given the probabilities of the observables we defined in the previous section we now
consider Shannon's entropy of these data. 
Shannon entropy was introduced by Claude Shannon \cite{s}.
See also \cite{ad} for a classic textbook on the subject and \cite{jb} for a recent approach on
the change in entropy associated with a measure-preserving function.

\begin{definition}
Let a set of possible events have probabilities $p_1, p_2, \dots , p_n$ such that 
$p_j >0$ for all $j=1, \dots, n$ and $\sum_{j=1}^n \ p_j = 1.$ The Shannon entropy is the function
on $(p_1, p_2, \dots, p_n)$ given by 

\begin{equation}
H_{n}(p_1, p_2, \dots, p_n)= - \sum_{j=1}^n p_j \ln(p_j) 
\end{equation}
\end{definition}
Note that naturally the Shannon entropy is in fact a sequence of functions on finite sets
of probabilities of events.\footnote{Here we only need to centre our attention in non-zero probability.}  
 
 The Shannon entropy is a unique function with certain properties which can be recalled in
 \cite{s} and \cite{ad}. 
 The one we consider here is that Shannon's entropy is maximal when the set of all probabilities 
 $(p_1, p_2, \dots, p_n)$ are equal. There is the inequality 
 
 \begin{equation}
H_{n}(p_1, p_2, \dots, p_n) \leq H_{n} \bigg( \frac{1}{n},  \frac{1}{n}, \dots, \frac{1}{n} \bigg) = \ln(n) 
\end{equation}

\subsection{One edge} If we consider the observable $\mathcal{O}$
consisting of a single edge labelled $j$, we have from our previous section
the set of probabilities corresponding to each spin, that is,
$(p_0, p_1, p_2 \dots, p_n)$, where $n=(r-1)$. The Shannon entropy of this 
example can be computed from formula $(10)$ and applying formula $(11)$ it is obviously bounded by

\begin{figure}[h]
\begin{center}
\includegraphics[width=0.9\textwidth]{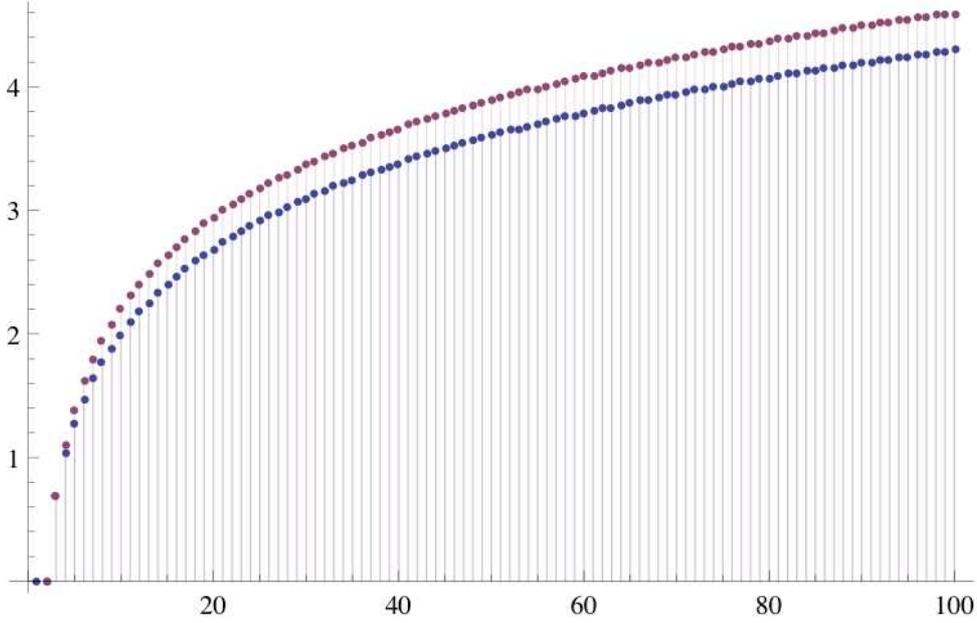}
\caption{The blue graph is the Shannon entropy function of the edge observable while the purple
one is $\ln(r-1)$. The abscissa axis represents different values of $r$.}
\end{center}
\end{figure}

 \begin{equation}
H_{(r-1)}(p_0, p_1, \dots, p_{(r-2)}) \leq \ln(r-1) 
\end{equation}
In Figure 1 we can see a discrete plot of the Shannon entropy function of the edge observable for different values of $r$. The blue one represents the Shannon entropy function and it is compared
against the growing of $\ln(r-1)$ which is drawn in purple.

Let us mention something about the Shannon entropy of the one edge observable.
Formula $(12)$ is an inequality which means that the major contribution to the entropy
of the one edge observable is given by $\ln(r-1)$ which is when all probabilities
are equal. However, not all probabilities are equal. According to previous section, 
the one edge observable probabilities are interpreted as measuring the area of 2-spheres
of different radii in $S^3$ which radius is $r/2\pi$.

It can be observed from formula $(8)$ that the most probable outcomes of geometrical 
measurements of the one edge observable are given by 2-spheres of
larger area. For large $r$ this happens when $j$ is between $(r-2)/2$ and $(r+2)/2$.

Once again; the entropy is dominated by the most probable outcomes, which in this case is given
by larger areas of 2-spheres in $S^3$. This is what we can say about this geometrical
measurements.

\subsection{The triangle}
 
If we consider the observable $\mathcal{O}$
consisting of the triangle labelled $i$, $j$ and $k$, we have from equation $(9)$
a set of probabilities 

\begin{equation}
\bigg\{ P_{(i,j,k)} = \frac{\dim_{q} i \dim_{q} j \dim_{q} k}{N^2} \ N_{i,j,k} \bigg\} 
\end{equation}
which runs over all different admissible colourings of a triangle. 
The Shannon entropy function is computed from formula $(10)$, and the following is true:

\begin{theorem}
The Shannon entropy of the triangle observable with the set of probabilities as in (13) is 
bounded by its maximal possible value

\begin{equation}
H_{n}(\{ P_{(i,j,k)} \}) \leq \ln\bigg( \frac{(r-1)\ r \ (r+1)}{6} \bigg) 
\end{equation}
which means that the number of non-zero probabilities is $n=(r-1) \ r \ (r+1)/6$.
\begin{proof}
It suffices to prove that the number of different admissible colourings\footnote{The reader may have already noticed that this has to do
with the dimension of the space of conformal blocks \cite{v}. Putting this aside for now we continue with our proof which is completely combinatorial.} 
of the triangle(equivalently of the theta
graph) equals $n=(r-1) \ r \ (r+1)/6$.

The set of colours is given by $L= \{ 0,1,2,...(r-2) \}$ and the admissibility conditions are given 
in equation $(2)$. 

By pondering a bit; it is noticed that in order for a triple $(i,j,k)$ to satisfy the 
triangle inequalities, if $i + j + k = 2s $
with $s \leq r-2$ then $i, j, k \in \{  0,1,2,...s \}.$ Once noticing this fact we need to count the number
of ways the following equation is satisfied

\begin{equation}
i + j + k = 2s \nonumber 
\end{equation}
By the inclusion-exclusion theorem\footnote{See appendix}, we first let the objects of
the set to be the number of ways
of decomposing an even number in a sum of three non-negative integers, which is given by

\begin{equation}
N= \left(
\begin{array}{c}
2s + 3 -1 \\
3-1 \\
  \end{array}
\right) = \left( \begin{array}{c}
2s + 2 \\
2 \\
  \end{array}
\right) = \frac{(2s+ 2)!}{2s!  \ 2} = \frac{(2s+1)(2s+2)}{2} \nonumber
\end{equation}
Let a solution have property $i$, if $i > s$.
Applying the inclusion-exclusion principle this means that

\begin{equation}
i \geq s+1 \ \Rightarrow \nonumber
\end{equation} 

\begin{equation}
(i + s+1) + j + k = 2s \ \ \ \Rightarrow \ \ \ i + j + k = s-1 \nonumber 
\end{equation}
Therefore 

\begin{equation}
N(i) = \left(
\begin{array}{c}
s+1 \\
2 \\
  \end{array}
\right) =
\frac{(s+ 1)!}{(s-1)!  \ 2} = \frac{s(s+1)}{2} \nonumber
\end{equation}
For $j$ and $k$ we equally have that $N(i)=N(j)=N(k).$
By the inclusion-exclusion principle 

\begin{equation}
N(i' \ j' \ k')= \frac{(2s+1)(2s+2)}{2} -  \frac{3s(s+1)}{2} = \frac{(s+1)(s+2)}{2} \nonumber
\end{equation}
$N(i' \ j' \ k')$ is the number of solutions in non-negative integers to the equation
$i + j + k = 2s$ for a particular value $s$.

Finally the number $n$ is given by summing over all $s$ 

\begin{equation}
n= \sum_{s=0}^{r-2} \frac{(s+1)(s+2)}{2} =  \frac{(r-1)\ r \ (r+1)}{6} \nonumber
\end{equation}
\end{proof}
\end{theorem}

It is easily seen that the number of different admissible colourings of the triangle is equivalent
to the number of different admissible colourings of the theta spin network graph. 

In a similar spirit to the one edge observable entropy for the triangle entropy we can say that  
the entropy is dominated by the
most probable geometrical outcomes, which happens for triangles with larger length sides. 
This happens for equilateral triangles. 

\bigskip

In this section we have studied the Shannon entropy of some spin network observables.
The outcomes of measurements of the observables are geometrical in nature, they are
mathematical and should only be seen in this way. A physical interpretation is less possible
since the theory we are dealing with here is the Turaev-Viro one. It is only a toy and simple
spin foam model of 3-dimensional quantum gravity. However let us only say that the measurements
we are talking about here, since they are geometrical, physically we could say that there is an 
apparatus which measures the geometry of space-time; and which for instance in the edge 
observable case it gives squared length measures with certain
probabilities. At the moment we are only proposing that this idea may be extended to 
the 4-dimensional case and which may give information about measures of the space-time geometry.

\bigskip

In the next section we generalise to further and general examples of spin network observables.

\section{Generalisations}

We can continue computing the Shannon entropy of different observable graphs. 
All these observables are imbedded on the triangulated three dimensional manifold $M$ 
as a subset of edges. 

It is important to point out that a very different Shannon entropy for the quantum 
gravitational field 
was studied in \cite{rv}. The entropy introduced in such reference is related to the 
Hamiltonian density matrix of loop quantum gravity and it is not known how 
to interpret it  in the language of a probability distribution.

In the present paper we have that the observable spin networks we are dealing with,
have a probability
distribution related to the colours of its edges. 
We now suggest that in a general way we can define the Shannon entropy associated to any 
spin network graph. Consider a trivalent graph with the following
properties:

Let $\Gamma_g$ be a trivalent graph with $2g-2$ vertices and $3g-3$ edges 
and with $g$ loops($g \geq 2$). 

Define a colouring of its edges from the set $L= \{ 0,1,2,...(r-2) \}$
to be admissible if at each vertex 
the colours of its three edges are admissible, that is, for each vertex we have that 
its three edges adjacent to it satisfy equation $(2)$.

The linear space generated by the admissible colourings of the graph 
$\Gamma_g$ is called the space of conformal blocks on a Riemann surface of genus 
$g$. (See \cite{tk}, \cite{sap}).  We denote this space by $V_{(r-2)}\Sigma_g$.
 
Therefore the inequality of theorem 1 suggests that the Shannon entropy of the triangle observable
(or theta spin network graph) is bounded by the logarithm of the 
dimension of the space of conformal blocks of the genus two torus of level $k=(r-2)$. 

The dimension of the space of conformal blocks $V_g$ on a genus $g$ 
Riemann surface was obtained in \cite{v}.
We can write down the inequality of theorem 1 as

\begin{equation}
H_{\dim V_{(r-2)} \Sigma_2}(\{ P_{(i,j,k)} \}) \leq \ln ( \dim V_{(r-2)} \Sigma_2 ) 
\end{equation}

\bigskip

\bigskip

\begin{theorem}
The Shannon entropy of a trivalent graph $\Gamma_g$  with $2g-2$ vertices and 
$3g-3$ edges and with $g$ loops($g \geq 2$)
is bounded by its maximal possible value
\begin{equation}
H_{n}(\{{\Gamma_g} _{(j_1, j_2, \cdots j_{3g-3})} \}) \leq \ln ( \dim V_{(r-2)} \Sigma_g ) 
\end{equation}
where $\{{\Gamma_g} _{(j_1, j_2, \cdots j_{3g-3})} \}$ is its set of probabilities and  $n=\dim V_{(r-2)} \Sigma_g$
is the number of non-zero probabilities.
\begin{proof}
The dimension of the space of conformal blocks is generated by the
number of admissible colourings of the corresponding spin network graph, therefore the inequality follows. 
\end{proof}
\end{theorem}
The difference to proving theorem 1 and theorem 2 is that in theorem 1,  we 
know exactly that the number of different admissible colourings of the theta
graph is given by $n= (r-1)\ r \ (r+1)/ 6$ since we computed it using
the inclusion-exclusion theorem. 

However, in theorem 2 we know that the number
of admissible colourings of the corresponding trivalent spin network graphs is given
by the dimension of the space of conformal graphs of the associated Riemann surface.
What we lacked of is the fact that we are not able to compute the number of 
admissible colourings using the same procedure as in theorem 1. Trying to apply
the same procedure is more challenging.
However the inequality still follows.

Now the reader may wonder, how exactly do we assign probabilities to each labelling of the graph $\Gamma_g$ \ ? 

The answer is as follows:  
Given a trivalent graph $\Gamma_g$  with $2g-2$ vertices and 
$3g-3$ edges and with $g$ loops($g \geq 2$)
the set of probabilities is calculated by

\begin{equation}
(\{ P_{(j_1, j_2,..., j_{3g-3})} \}) \sim \bigg< \Gamma_{g}(j_1, j_2,..., j_{3g-3})  \bigg>^{2}
\end{equation}
where the squared evaluation of the graph on the right hand side of the formula $(17)$ is calculated
using the recoupling theory of the spin network graph.

For instance for the case of genus $g=3$ the spin network graph is given by a tetrahedron.
Its set of probabilities implies the bound inequality 

\begin{equation}
H_{\dim V_{(r-2)} \Sigma_3}(\{ P_{(j_1, j_2, j_3, j_4, j_5, j_6)} \}) \leq \ln ( \dim V_{(r-2)} \Sigma_3 )
\end{equation}
We have defined the Shannon entropy of spin network graph observables. We have shown that the 
value of the entropy is bounded. A precise meaning for the measures of the observables were made
for the one edge observable and the triangle observable. An interpretation of the general
spin network measurements and of the Shannon entropy is a bit unknown at the moment by the author.
It can only be said that something geometrical about the space-time is to be expected.
Maybe the geometry of space-time can be recovered from the information of the measurements 
we have described, and it can be indeed studied for 4-dimensional quantum gravity.
However for the moment we do not know about it and what we have presented are
good mathematical ideas to be interpreted in future.

\section{Conclusions}

We have introduced a way to define the Shannon entropy of spin network graphs 
in the case of Euclidean
three dimensional quantum gravity. 

When considering the set of colours $L= \{ 0,1,2,...(r-2) \}$,
we have written down an inequality which bounds the Shannon entropy of a spin network graph
$\Gamma_g$.
In particular we have proved that
the Shannon entropy of a triangle(or theta spin network graph)
is bounded by $n=(r-1) \ r \ (r+1)/6$; and we have proved it in a completely combinatorial
way.   

Now, we pose the following questions for a future work. First of all, what is the Shannon entropy
of a spin network graph? Or in other words, what information can we extract from it?
How is it related to the entropy studied in \cite{gi2}, \cite{gi3}? 

In the combinatorial context an interesting question is; can we prove for the 
case of the tetrahedron that the Shannon entropy is bounded by the logarithm of 
$n= (r-1) \ r^2 \ (r+1)/180$. That is; is there a proof analogous to the one of theorem 1 
which uses the
the Inclusion-Exclusion theorem ?

For instance, in \cite{v}, the dimension of the space of conformal blocks
has an analytic expression for any surface of any genus $g$. 
 
Our question now is; is there a way to prove combinatorially(using the Inclusion-Exclusion technique)
that the dimension of 
the space conformal blocks of the Riemann surface of genus $3$ is the account number of 
formula (16). And if such combinatorial technique works for the genus $2$ and genus $3$ surfaces,
can we find a combinatorial algorithm(using the Inclusion-Exclusion technique) to produce
a formula for the dimension of the space of conformal blocks other than the analytical one
found in \cite{v} and in all of the literature?

Prove that for the tetrahedron graph which represents
a Riemann surface of genus $g=3$, 
$\dim V_{(r-2)} \Sigma_3 \sim (r-1) \ r^2 \ (r+1)/180$ using only combinatorial techniques
(The Inclusion-Exclusion theorem.)

Another very interesting question is how the Shannon entropy we studied here
is related to the one studied in \cite{rv}.

\bigskip

\bigskip 

\bigskip

\bigskip

\danger{Acknowledgements}: Firstly I want to thank so deeply to Alejandro Perez and to Carlo Rovelli
for their suggestion on the subject of this work. I also thank them for their
hospitality at the Centre de Physique Th\'eorique de Luminy in Marseille, France.
I also thank Francesco Costantino from Institut de Recherche Math\'ematique Avanc\'ee,
for correspondence related to conformal blocks.

\appendix
\section{The Inclusion-Exclusion Principle}

Here we basically describe the inclusion-exclusion principle without proof and refer 
to \cite{bm} for a deeper understanding of the process. 

\bigskip

\danger{Theorem A.1.} 
\emph{The equation}
\begin{equation}
x_1 + x_2 + \cdots + x_n = m \nonumber
\end{equation}
\emph{has exactly} 
\begin{equation}
\left(
\begin{array}{c}
m+n-1 \\
n-1 \\
  \end{array}
\right) \nonumber
\end{equation}
\emph{solutions in non-negative integers.}

\bigskip

Consider a set of $N$ objects and some properties denoted by $j_1, j_2, \cdots j_n.$
\footnote{Note that for convenience we are using the same notation that resembles spins which label spin network graphs.} Some of the $N$ objects may have one, two or more of the properties
and some others none of them. The symbol $N(j_\mu \  j_\nu, \cdots j_\rho)$ denotes the number
of objects having properties $j_\mu \  j_\nu, \cdots j_\rho.$ 
$N(j'_\mu \  j'_\nu, \cdots j'_\rho)$ denotes the number of objects with none of the properties.

\bigskip

\danger{Theorem A.2.} \ \emph{(The Inclusion-Exclusion Principle)}
\begin{align}
N(j'_1 \  j'_1, \cdots j'_n)= \ & N - N(j_1) - N(j_2) - \cdots - N(j_n) \nonumber \\
                                               & + N(j_1 j_2) + N(j_1 j_3) + \cdots + N(j_{n-1}j_n) \nonumber \\
                                               & - N(j_1 j_2 j_3) - N(j_1 j_2 j_4) - \cdots - N(j_{n-2} j_{n-1}j_n) \nonumber \\
                                               & + \cdots \ \ \ \ \cdots \ \ \ \ \cdots  \nonumber \\
                                               & + (-1)^n N(j_1, j_2, \cdots j_n) \nonumber
\end{align}

\bigskip

This principle theorem is applied in section 3 for proving theorem 1 and it is an example of its use.

\end{document}